\documentclass[conference]{IEEEtran}
\IEEEoverridecommandlockouts

\usepackage{cite}
\usepackage{amsmath,amssymb,amsfonts}
\usepackage{algorithmic}
\usepackage{graphicx}
\usepackage{textcomp}
\usepackage{xcolor}
\usepackage{url}
\usepackage{booktabs}
\usepackage{array}
\usepackage{hyperref}
\hypersetup{
    colorlinks=true,
    linkcolor=black,
    citecolor=black,
    urlcolor=blue
}

\def\BibTeX{{\rm B\kern-.05em{\sc i\kern-.025em b}\kern-.08em
    T\kern-.1667em\lower.7ex\hbox{E}\kern-.125emX}}

\begin{document}

\title{Pre-Execution Query Slot-Time Prediction in Cloud Data Warehouses: A Feature-Scoped Machine Learning Approach}

\author{
\IEEEauthorblockN{Prashant Kumar Pathak}
\IEEEauthorblockA{\textit{Independent Researcher} \\
Santa Clara, CA, USA \\
prashant.pathak@ieee.org \\
ORCID: 0009-0008-8744-9025}
}

\maketitle

\begin{abstract}
Cloud data warehouses bill compute resources based on slot-time consumed. In shared multi-tenant analytics environments, query execution cost is highly variable and difficult to estimate before execution, leading to budget overruns, inefficient scheduling, and degraded tenant experience. Existing cost estimators based on static query planner heuristics fail to capture effects of complex SQL structure, data distribution skew, and workload contention. We present a feature-scoped machine learning approach that predicts BigQuery slot-time before execution using only pre-execution observable signals: a structured query complexity score derived from SQL operator costs, data volume features derived from planner estimates and workload metadata, and textual features from query text. We deliberately exclude runtime factors (concurrent slot-pool utilization, cache state at execution, realized data skew) that are unknowable at submission time. The model uses a HistGradientBoostingRegressor trained on log-transformed slot-time, with a TF-IDF + TruncatedSVD-512 text pipeline fused with numeric and categorical features. Trained on 749 queries across seven deployment environments of varying size and evaluated out-of-distribution on 746 queries from two held-out environments not used in training, the model achieves MAE 1.17 slot-minutes, RMSE 4.71, and 74\% explained variance on the full test workload. On the cost-significant query subset (slot-time $\geq$ 0.01 min, N=282) the model achieves MAE 3.10 versus 4.95 for a predict-mean baseline and 4.54 for a predict-median baseline, a 30--37\% MAE reduction. On long-tail queries ($\geq$ 20 min, N=22) the model does not outperform trivial baselines, consistent with the hypothesis that long-tail queries are dominated by unobserved runtime factors outside the current feature scope. A complexity-routed dual-model architecture is described as a practical refinement, and directions for closing the long-tail gap are identified as future work.
\end{abstract}


\section{Introduction}

Cloud data warehouses such as Google BigQuery, Amazon Redshift, and Snowflake have transformed large-scale analytics by decoupling compute from storage and enabling on-demand query execution over petabyte-scale datasets. This architecture introduces a billing model fundamentally different from traditional database systems: tenants are charged for compute resources their queries consume, measured in units such as slot-time, bytes processed, or credit consumption. This compute-based billing creates a critical operational problem---query cost is not known until after the query executes.

Engineers submitting analytical queries, automated reporting pipelines, and ad hoc workloads routinely produce unexpected cost spikes that exhaust monthly budgets, delay dependent jobs, and degrade platform experience for co-resident tenants. Existing cost estimation mechanisms provided by cloud data warehouse platforms rely on static query planner heuristics: row count estimates, join-order optimization, and simplified operator cost models. These approaches are systematically inaccurate under real-world conditions because they do not model complex SQL structure, data distribution skew, probabilistic cache effects, or multi-tenant contention. The result is that static estimates frequently differ from actual execution costs by one or more orders of magnitude \cite{neo}\cite{kipf}.

This paper investigates how far pre-execution slot-time prediction can be advanced using machine learning on features observable at query submission time. The key design choice is scope: we deliberately restrict the model to features computable before execution begins. This excludes powerful signals---concurrent slot-pool state, realized cache hit rates, runtime data skew---that would improve accuracy but are fundamentally unknowable at the moment a prediction is needed. The contribution is a characterization of what is predictable from pre-execution signals alone, what residual variance remains attributable to unobserved runtime factors, and what practical accuracy can be achieved with a gradient-boosting approach on engineered features.

\subsection{Contributions}

This paper makes the following contributions. First, we present an explicitly feature-scoped prediction pipeline combining a structured query complexity score, data volume features, and TF-IDF textual features over query text, fused through a column-transformer architecture with dimensionality reduction via TruncatedSVD. Second, we describe a structural query complexity metric calculated from a weighted tally of SQL operator costs, providing a single interpretable numeric feature that carries substantial predictive signal. Third, we evaluate a HistGradientBoostingRegressor on 746 real BigQuery queries drawn from a production-scale multi-tenant cloud analytics workload, using an out-of-distribution test split where the evaluation queries were drawn from deployment environments entirely excluded from training. Fourth, we report tiered evaluation results disclosing where the model succeeds (74\% explained variance full workload; 30--37\% MAE reduction over trivial baselines on cost-significant queries) and where it does not (long-tail queries $\geq$20 slot-min). Fifth, we describe a complexity-routed dual-model architecture as a practical refinement that improves accuracy by specializing models to different query complexity regimes.

\section{Background and Motivation}

\subsection{Cloud Data Warehouse Billing Models}

Cloud data warehouses employ fundamentally different billing models from traditional relational database deployments. In Google BigQuery's on-demand mode, queries are billed per byte of data scanned. In slot-reservation mode, tenants purchase dedicated compute capacity measured in slots---units of CPU, memory, and network bandwidth---and are billed for slot-time consumed. Amazon Redshift bills for provisioned node-hours; Snowflake bills for compute credit consumption proportional to virtual warehouse size and query duration. The common problem across these models is that the billing quantity is a function of query execution behavior that is not precisely known before execution.

\subsection{Factors Affecting Slot-Time}

Query slot-time in a distributed cloud data warehouse is a function of several factors, which we partition by whether they are observable at query submission time.

\textbf{Pre-execution observable factors} (used in this work):
\begin{itemize}
\item Query structure: join counts, aggregation operators, subquery presence, window functions, CTEs, UDF usage, and an aggregate complexity score
\item Data volume: bytes processed and billed estimates from the planner, workload-level cardinality metrics (account and resource counts), log-scaled volume transformations
\item Tenant and workload identifiers: project and dataset identifiers, asset-type categorical features, cache-hit indicator
\end{itemize}

\textbf{Runtime-only factors} (not used, require execution):
\begin{itemize}
\item Concurrent slot-pool utilization at execution time
\item Cache hit decisions made during execution
\item Realized data skew encountered during joins and aggregations
\item Slot scheduler decisions under contention
\item Intermediate shuffle volumes
\end{itemize}

This decomposition motivates the feature-scoped design. Pre-execution prediction has a principled upper bound on accuracy set by the variance contribution of runtime-only factors. Quantifying how close a learned model can come to that bound---and identifying which query classes are most affected by the gap---is the empirical contribution of this paper.

\subsection{Limitations of Static Cost Estimation}

Query planners in distributed data warehouses estimate execution cost using statistics-based cost models: table row counts, column cardinality estimates, predicate selectivity estimates, and operator cost coefficients. Complex SQL structures---nested subqueries, correlated subqueries, window functions, user-defined functions---create execution plan shapes that static cost models approximate poorly. Column statistics reflect global distributions but not partition-level skew. Cache hit rates are highly variable and depend on temporal locality. Our approach does not attempt to improve the planner's internal estimates; we build an external prediction layer trained on observed end-to-end outcomes.

\section{System Overview}

The prediction system comprises four components operating as a sequential pipeline: data collection from historical execution telemetry, feature extraction from raw SQL and metadata, model training via gradient-boosting regression on log-transformed targets, and inference producing pre-execution predictions for new queries.

\subsection{Data Sources}

Historical query execution data is sourced from BigQuery's \texttt{INFORMATION\_SCHEMA.JOBS} view, which exposes per-query execution records including raw SQL text, project and dataset identifiers, bytes processed and billed, total slot-milliseconds consumed, total elapsed duration, cache hit indicator, execution region, and creation timestamp. Supplementary workload metadata---tenant-level account and resource counts, asset-type catalogs---is joined to query records to enrich the feature set. This dataset is filtered to remove trivial DDL statements, queries with anomalous execution records reflecting system errors, and queries that reached the execution timeout (discussed further in Section~VII).

\subsection{Prediction Target}

The system predicts total slot-time consumed, expressed in minutes for interpretability. The target is log-transformed prior to training via $y = \log(1 + \text{slot\_time\_min})$ (\texttt{log1p}). Query slot-time spans multiple orders of magnitude---from fractional milliseconds for trivial queries to hours for full table scans---and log transformation stabilizes variance across this range, reducing the influence of extreme values on model training. Predictions are back-transformed to the original scale at inference time via \texttt{expm1}.

\section{Feature Extraction}

The feature extraction pipeline transforms raw SQL query text, planner estimates, and workload metadata into a structured feature vector for each query. Features fall into four categories corresponding to the pre-execution observable factors enumerated in Section~II.

\subsection{Query Complexity Score}

A central engineered feature is a structured complexity score that summarizes SQL operator cost. Each query's SQL text is parsed to count occurrences of structurally costly operators; each occurrence contributes a weight to the query's total complexity score. The weights were calibrated empirically by correlating operator presence with observed slot-time on the training corpus; the final weight table is shown in Table~\ref{tab:complexity}.

\begin{table}[htbp]
\caption{Query Complexity Scoring Weights}
\label{tab:complexity}
\centering
\small
\begin{tabular}{@{}lr lr@{}}
\toprule
\textbf{Operator} & \textbf{Weight} & \textbf{Operator} & \textbf{Weight} \\
\midrule
Join & 3 & Unnest & 2 \\
Cross Join & 5 & Merge & 4 \\
Group By & 2 & Update & 3 \\
Distinct & 2 & Insert & 1 \\
Order By & 2 & WITH CTE & 1 \\
Window Function (OVER) & 3 & Subselect & 2 \\
Regex Function & 4 & Array/Struct & 1 \\
SQL UDF & 1 & Having & 1 \\
JS UDF & 6 & & \\
\bottomrule
\end{tabular}
\end{table}

The total complexity score is the sum of (count $\times$ weight) across all operator types. For example, a query with two GROUP BY and two DISTINCT contributes $(2 \times 2) + (2 \times 2) = 8$ to the complexity score. JavaScript UDFs carry the highest per-occurrence weight (6) reflecting their outsized impact on execution cost due to cross-runtime context switches; Cross Join carries weight 5 because of its quadratic cardinality behavior. The complexity score is used both as a standalone numeric feature and as a routing signal for the dual-model architecture described in Section~V.

\subsection{Data Volume Features}

Volume features capture the scale of data the query will process:
\begin{itemize}
\item \textit{Bytes processed and billed}: available from the query planner before execution
\item \textit{Workload cardinality}: total tenant account count, total resource count
\item \textit{Provider-partitioned accounts}: number of accounts per cloud provider (AWS, GCP, Azure)
\item \textit{Asset-type counts}: counts of data entities of relevant types present in the query scope
\item \textit{Derived features}: log1p-transformed bytes, bytes-per-account, bytes-per-resource ratios
\end{itemize}

Volume features are log-scaled to match the log-transformed prediction target and to reduce leverage of extreme values.

\subsection{Textual Features}

Raw query text is vectorized using TF-IDF over n-grams covering unigrams and bigrams. Bigrams capture local operator sequences that are predictive of execution behavior---for example, the bigram \texttt{GROUP BY} carries more signal than the tokens in isolation. Query text is cleaned to strip project-qualified table references, normalize whitespace, and uppercase SQL keywords before tokenization. String literals and numeric constants are replaced with placeholder tokens to reduce vocabulary fragmentation from parameterized queries that differ only in literal values.

The high-dimensional sparse TF-IDF output is reduced to a 512-dimensional dense representation via Truncated Singular Value Decomposition (SVD). This reduction serves two purposes: it produces a dense feature block compatible with gradient-boosting algorithms that do not natively handle sparse inputs, and it concentrates the textual signal into a tractable dimensionality for gradient-boosted trees.

\subsection{Categorical Features}

Categorical features include project and dataset identifiers, asset-type identifiers (top-N most frequent types, one-hot encoded), the binary cache-hit indicator, and execution region. Categorical identifiers carry implicit workload signal that query text alone does not capture---different projects exhibit systematically different query patterns and data volumes.

\section{Model Architecture and Training}

\subsection{Feature Fusion Pipeline}

The four feature categories are fused using a column-transformer pipeline with category-appropriate transformations applied in parallel:
\begin{itemize}
\item Text: TF-IDF (1--2 grams) $\to$ TruncatedSVD (512 components) $\to$ dense
\item Numerics: StandardScaler applied to complexity score, account counts, resource counts, bytes metrics
\item Volumetric: log1p-transformed bytes, bytes-per-account, bytes-per-resource
\item Categorical: One-hot encoding of asset-type identifiers and provider flags $\to$ dense
\end{itemize}
Outputs are concatenated into a single dense feature matrix suitable for gradient-boosted tree training.

\subsection{Regression Model}

The prediction system uses a \textbf{HistGradientBoostingRegressor} (histogram-based gradient boosting over decision trees) with learning rate approximately 0.07 and 300 boosting iterations. Gradient boosting was selected after comparative evaluation against linear baselines (Ridge with $\ell_2$ regularization, ElasticNet, Huber regression) and an SGDRegressor with Huber loss. Ridge provided a strong baseline and remained interpretable but underperformed gradient boosting on the heavy-tailed target. ElasticNet underperformed Ridge while taking substantially longer to fit. Huber-based linear models were explored but did not beat Ridge on our evaluation. Gradient boosting's ability to capture non-linear interactions among the engineered features---particularly between complexity score, volume, and text-derived signals---produced the best out-of-distribution generalization in our experiments.

\subsection{Complexity-Routed Dual-Model Architecture}

During model development we observed that queries above a complexity-score threshold exhibit qualitatively different cost behavior: their slot-time is more sensitive to structural operators and less sensitive to volume signals, compared to simpler queries where volume dominates. We therefore use a complexity-routed architecture in which queries are routed to one of two models based on complexity score:
\begin{itemize}
\item \textit{Simple-query model}: trained and applied to queries with complexity score $<$ 26
\item \textit{Complex-query model}: trained and applied to queries with complexity score $\geq$ 26
\end{itemize}
Both sub-models share the same pipeline and algorithm (HistGradientBoostingRegressor on the fused dense feature matrix); the split provides each sub-model with a more uniform training distribution, improving accuracy over a single-model baseline by reducing within-model variance in the complexity dimension.

\subsection{Training Procedure}

The pipeline is fit on the training corpus, producing a serialized artifact bundle (column transformer, TF-IDF vocabulary, SVD components, scaler statistics, trained regressors, and metadata) using joblib. Hyperparameters---learning rate, number of iterations, regularization---are tuned via cross-validation on the training split. At inference, the same pipeline is loaded and applied to new queries, producing log-space predictions that are back-transformed via \texttt{expm1}.

\section{Evaluation}

\subsection{Experimental Setup}

We evaluate on real BigQuery query execution logs drawn from a production multi-tenant cloud analytics workload. The training corpus comprises 749 queries executed across seven deployment environments of varying size (very small, small, medium, and large), each with different data shapes and cardinality profiles. The held-out test set consists of 746 queries drawn from two environments that were entirely excluded from training---an out-of-distribution evaluation where the model must generalize to workloads whose data shape it has never seen.

The test environments represent a large-scale operational workload, with tenant-level cardinality in the tens of thousands of accounts and tens of millions of resource records. The test set spans a wide dynamic range of query slot-time: from sub-millisecond slot-time for trivial metadata queries up to approximately 125 slot-minutes for complex high-cardinality joins, reflecting realistic workload skew. The distribution is heavily tailed: median slot-time 0.0005 minutes (typical of metadata and cache-hit queries), mean 4.5 minutes, 75th percentile 10.2 minutes, 95th percentile 14.4 minutes.

Queries that reached the execution timeout were excluded from both training and testing. Timeout queries have slot-time truncated at the timeout boundary rather than reflecting their intrinsic cost, and including them would contaminate both the training signal and the evaluation metrics. Identifying timeout-prone queries before execution is itself a useful problem and is identified as future work.

\subsection{Overall Performance}

On the complete 746-query held-out test set, the model achieves:
\begin{itemize}
\item Mean Absolute Error: 1.17 slot-minutes
\item Root Mean Squared Error: 4.71 slot-minutes
\item Explained variance: 0.74
\item Prediction variance / actual variance ratio: 0.47
\end{itemize}

\begin{figure}[htbp]
\centerline{\includegraphics[width=\columnwidth]{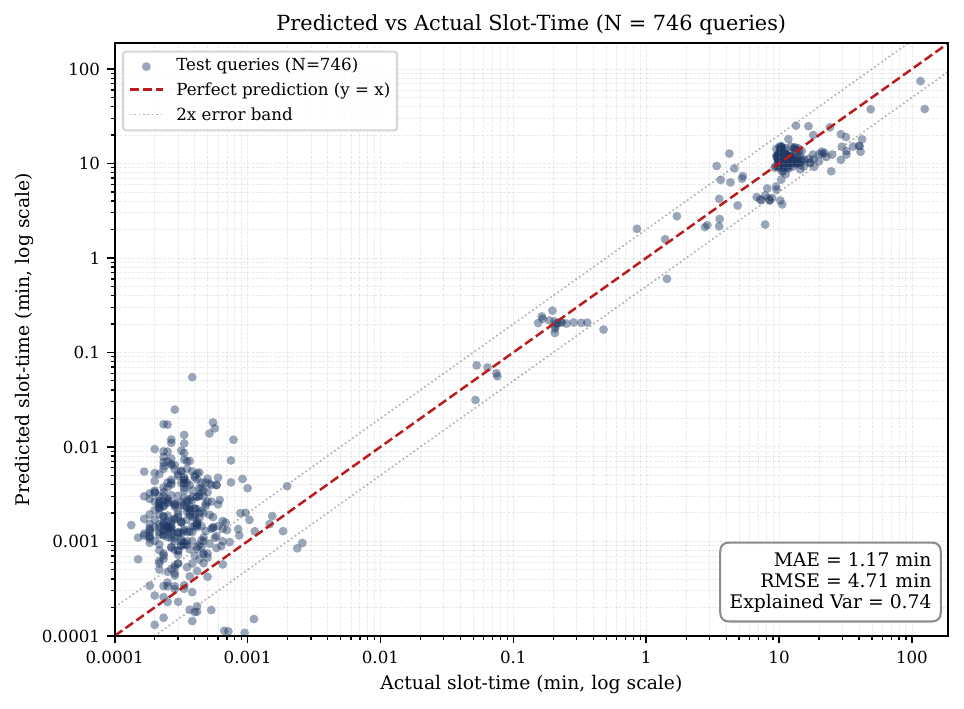}}
\caption{Predicted versus actual slot-time for 746 held-out test queries (log-log scale). The dashed red line represents perfect prediction. Dotted lines indicate the 2$\times$ error band. The majority of predictions fall within a factor of two of the actual value across four orders of magnitude.}
\label{fig:scatter}
\end{figure}

Figure~\ref{fig:scatter} shows predicted versus actual slot-time on a log-log scale across the full dynamic range. The majority of points cluster within the 2$\times$ error band, indicating consistent tracking of actual slot-time across four orders of magnitude. The variance ratio of 0.47 indicates predictions are somewhat conservative on the extreme tail---the model regresses toward the mean on rare high-cost queries, a behavior consistent with the heavy-tailed training distribution and the exclusion of runtime factors that drive extreme costs.

\subsection{Tiered Evaluation and Baseline Comparison}

\begin{figure}[htbp]
\centerline{\includegraphics[width=\columnwidth]{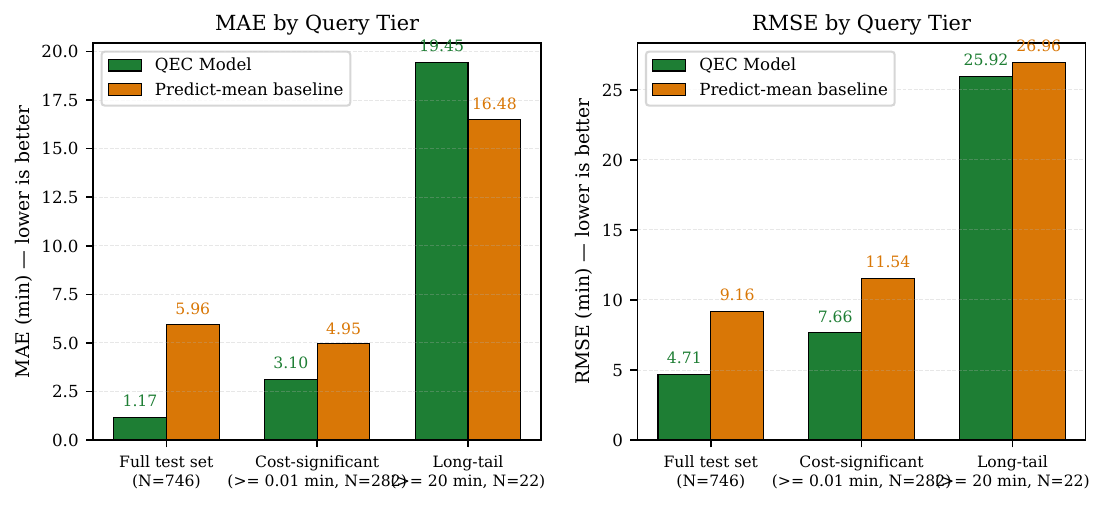}}
\caption{Tiered evaluation across the full test set, cost-significant subset, and long-tail subset. The model beats the predict-mean baseline on the first two tiers but loses on the long-tail tier, where unobserved runtime factors dominate.}
\label{fig:tiered}
\end{figure}

Aggregate metrics on heavy-tailed workloads can obscure where a model contributes value and where it does not. We partition the test set into three operationally meaningful tiers and compare the model against a predict-mean baseline (which predicts the test-set mean for every query). Results are shown in Table~\ref{tab:tiered} and Figure~\ref{fig:tiered}.

\begin{table}[htbp]
\caption{Tiered Evaluation: Model versus Predict-Mean Baseline}
\label{tab:tiered}
\centering
\small
\begin{tabular}{@{}lrrrr@{}}
\toprule
\textbf{Tier} & \textbf{N} & \textbf{Model MAE} & \textbf{Baseline MAE} & \textbf{Reduction} \\
\midrule
Full test set & 746 & 1.17 & 5.96 & 80\% \\
Cost-significant & 282 & 3.10 & 4.95 & 37\% \\
Long-tail ($\geq$20 min) & 22 & 19.45 & 16.48 & $-18$\% \\
\bottomrule
\end{tabular}
\vspace{2pt}

\footnotesize Cost-significant tier: queries with actual slot-time $\geq$ 0.01 min. Units: slot-minutes.
\end{table}

\textbf{Full test set.} The model achieves MAE 1.17 versus baseline 5.96, an 80\% reduction. This aggregate result is partially driven by a large population of trivial queries (N=464, slot-time $<$ 0.01 min---typically metadata or cache-hit queries) on which the model correctly predicts near-zero values.

\textbf{Cost-significant queries (N=282).} This tier represents queries with measurable compute cost where pre-execution cost prediction is operationally relevant---queries a FinOps team might want advance warning about. The model achieves MAE 3.10 versus 4.95 for predict-mean and 4.54 for predict-median, a 30--37\% reduction over trivial baselines. This is the tier where the model demonstrates its primary practical contribution.

\begin{figure}[htbp]
\centerline{\includegraphics[width=\columnwidth]{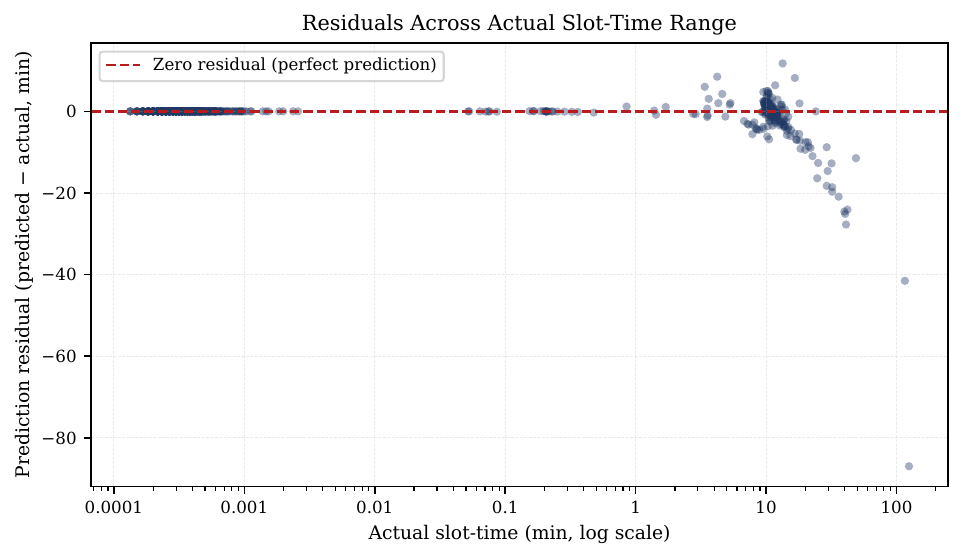}}
\caption{Prediction residuals across the actual slot-time range. Systematic negative residuals at high actual slot-time ($>$10 min) indicate the model under-predicts for long-tail queries, consistent with the hypothesis that unobserved runtime factors dominate in this regime.}
\label{fig:residuals}
\end{figure}

\textbf{Long-tail queries (N=22, slot-time $\geq$ 20 min).} On this subset, the model's MAE (19.45) is higher than the predict-mean baseline's MAE (16.48). The model systematically under-predicts for queries in this tier, visible in Figure~\ref{fig:residuals} as a negative-residual pattern at high actual slot-time. This is consistent with the feature-scope limitation: long-tail queries are those most affected by unobserved runtime factors---slot-pool contention, realized data skew, cache misses on large working sets---which the current feature set cannot capture. The complexity-routing architecture partially mitigates but does not eliminate this gap. Quantifying the predictability gap attributable to runtime factors, and investigating whether additional feature families can close it, is a primary direction for future work.

\section{Threats to Validity}

\textbf{Internal validity.} The baseline comparison uses test-set mean rather than strict training-set mean due to data provenance constraints at analysis time; this introduces a minor optimistic bias favoring the baseline. The tier boundaries (0.01 min and 20 min) were chosen based on inspection of the distribution; results are stable to small perturbations of these boundaries. The complexity-routing threshold (26) was tuned on the training set; its optimal value may differ for substantially different workloads.

\textbf{External validity.} Evaluation is on 746 queries from a production multi-tenant cloud analytics workload of a specific shape. The feature pipeline requires only standard BigQuery \texttt{INFORMATION\_SCHEMA} access and portable SQL parsing, but generalization to substantially different workloads---different industries, different query composition patterns, different cardinality profiles---would require separate evaluation. The long-tail performance gap observed here is expected to generalize: any pre-execution model using the same feature family will face the same fundamental limitation on runtime-dominated queries.

\textbf{Feature-scope limitations.} By construction, this model does not use runtime-only features. The residual error---particularly on long-tail queries---therefore reflects the principled upper bound of what pre-execution prediction can achieve with this feature family, not a failure of the learning procedure. A fair comparison against runtime-aware baselines would require an evaluation framework beyond the scope of this paper.

\textbf{Timeout exclusion.} Queries that timed out were excluded from training and evaluation. A production deployment would need to either handle timeout prediction as a separate classification problem or treat high-predicted-cost queries conservatively until a timeout model is available.

\section{Proposed Integration Architecture}

Although the evaluation is a held-out research evaluation rather than a production deployment, the prediction pipeline is designed with three natural production integration points in mind.

\subsection{Pre-Submission Query Advisor}

The prediction pipeline would be invoked before a query is dispatched to the warehouse execution engine. If predicted cost exceeds a configurable threshold, the advisor surfaces a warning to the submitter. For automated pipelines, queries above a hard cost threshold could be held for review or rerouted to a lower-cost execution environment. Feature extraction and model inference on the serialized pipeline complete in well under 100 ms for typical query lengths, indicating prediction overhead acceptable for an interactive advisor.

\subsection{Dynamic Slot Scheduler}

In slot-reservation deployments, pre-execution cost predictions could be passed to the slot scheduler as priority hints. Low-cost queries dispatched immediately; high-cost queries scheduled during off-peak windows. Given the observed long-tail limitation, a production integration should flag high-predicted-cost queries as high-uncertainty rather than taking irreversible scheduling actions on point predictions alone.

\subsection{Cost Forecast Dashboard}

For scheduled reporting pipelines and batch analytics jobs, the prediction pipeline could generate periodic cost forecasts by running against scheduled query sets. These forecasts, surfaced alongside historical actuals, would enable analytics engineers and FinOps teams to identify high-cost jobs before execution and intervene through query optimization, scheduling changes, or resource allocation adjustments.

\section{Related Work}

Query cost estimation in relational databases has a long history. Selinger et al.~\cite{selinger} introduced the System R cost model using table statistics and operator cost coefficients for join ordering. Our work differs in that we do not modify or extend the query planner---we treat the execution engine as a black box and learn cost from observed execution telemetry.

Learned query optimization has attracted significant recent interest. Marcus et al.~\cite{neo} propose Neo, a learned query optimizer using tree-structured neural networks over query plan representations; Bao~\cite{bao} makes learned query optimization practical for real workloads. Kipf et al.~\cite{kipf} demonstrate that learned cardinality estimators significantly outperform traditional approaches on correlated joins. Our approach is complementary: rather than improving internal planner estimates, we build an external prediction layer trained on end-to-end execution outcomes and restrict to pre-execution observable features.

Gradient-boosted tree models for tabular prediction problems have been extensively studied~\cite{gbm}; their strong performance on heterogeneous feature sets with non-linear interactions motivated our choice of HistGradientBoostingRegressor for this application. Dimensionality reduction via Truncated SVD on TF-IDF features is a standard text-processing pattern adapted here to produce dense features compatible with tree-based models.

Cloud query workload management has been studied for Redshift~\cite{redshift}, BigQuery~\cite{dremel}, and Snowflake~\cite{snowflake}. These systems implement workload management policies based on query priority, resource class assignment, and admission control. Pre-execution prediction provides a cost signal that can drive more precise admission control and scheduling decisions than rule-based assignment.

Multi-tenant database performance under shared buffer pools has been studied~\cite{aurora}\cite{multitenant}, with emphasis on isolation and per-tenant caching. The long-tail performance gap we observe is consistent with findings in these studies that shared-resource contention contributes substantial unpredictability to query latency.

\section{Discussion and Future Work}

\subsection{Generalization Across Warehouses}

While developed in the context of BigQuery, the approach generalizes to other cloud data warehouses. The feature extraction pipeline requires only standard SQL parsing and is not BigQuery-specific. Training data sources are available in Redshift (\texttt{STL\_QUERY}), Snowflake (\texttt{QUERY\_HISTORY}), and other platforms. Billing units differ across platforms, but the prediction target can be adapted: bytes billed for on-demand BigQuery, credits for Snowflake, node-hour fractions for Redshift.

\subsection{Key Engineering Lessons}

Several lessons emerged from iterative model development. First, data and feature engineering had substantially larger impact on prediction accuracy than model algorithm selection: the same HistGradientBoostingRegressor applied to a richer feature set produced far better results than any linear model on a sparser feature set. Second, many iterations produced only small gains---the 30--37\% MAE reduction over baselines was achieved through many small feature improvements rather than any single architectural breakthrough. Third, preventing data leakage between training and test splits required careful attention because deployment environments in the training pool share schema patterns with held-out environments; strict environment-level splitting was necessary for honest out-of-distribution evaluation. Fourth, complexity-based routing produced meaningful gains where a unified model plateaued.

\subsection{Closing the Long-Tail Gap}

The primary direction for strengthening this work is reducing the long-tail prediction gap. Three feature-family extensions appear most promising:

\textbf{Data distribution features.} Partition-level cardinality, join-key skew estimates derivable from histograms, and per-table update frequency could capture factors currently hidden in runtime-only variance.

\textbf{Timeout classification.} A companion classifier predicting whether a query will reach the execution timeout would allow the cost regressor to defer on high-risk queries rather than silently under-predict.

\textbf{Partial runtime observability.} Current slot-pool utilization and recent query history are observable at submission time and could be added as features, moving the model from strictly pre-execution to submission-time prediction with limited runtime awareness.

\subsection{Uncertainty Quantification}

The current model produces point predictions only. Given the observed long-tail gap, adding uncertainty estimation via quantile regression or conformal prediction intervals would enable a production integration to respond appropriately to prediction confidence---flagging high-uncertainty predictions for human review rather than issuing potentially misleading point estimates.

\subsection{Model Drift and Retraining}

Query execution cost is influenced by factors that evolve over time: data volumes grow, query patterns shift, and platform software changes. A static model will exhibit drift as the distribution of incoming queries diverges from the training distribution. Periodic retraining on a rolling window of recent execution records, triggered by monitoring of prediction error drift, maintains accuracy without manual intervention.

\section{Conclusion}

We presented a feature-scoped machine learning approach to pre-execution query slot-time prediction in cloud data warehouses, combining a structured complexity score, data volume features, and TF-IDF textual features fused through a dense pipeline and modeled with HistGradientBoostingRegressor. By explicitly enumerating factors affecting slot-time and restricting the model to pre-execution observable features, we produce a system whose capabilities and limitations are principled rather than accidental.

Evaluated on 746 real BigQuery queries in an out-of-distribution test drawn from two deployment environments not used in training, the model achieves MAE 1.17 slot-minutes and 74\% explained variance on the full workload, and a 30--37\% MAE reduction over trivial baselines on cost-significant queries---the tier where pre-execution prediction carries operational value. A complexity-routed dual-model architecture improves accuracy by specializing sub-models to different query complexity regimes. Performance on long-tail queries ($\geq$20 slot-min) does not exceed trivial baselines, consistent with the hypothesis that these queries are dominated by unobserved runtime factors that no purely pre-execution model can capture.

The broader contribution is a framework for honest evaluation of pre-execution prediction: identify the feature scope, report tiered results, and attribute residual error to its true cause. Machine learning on cloud execution telemetry has real potential to transform cost management from reactive post-execution analysis to proactive pre-execution intelligence---but only when scope is honestly characterized and the boundary between predictable and unpredictable is respected.

\end{document}